\documentclass[pra,twocolumn,superscriptaddress,notitlepage,fixfloat]{revtex4-1}
\usepackage{graphicx,amsmath,amsfonts,amssymb,upgreek,txfonts,color}
\usepackage[colorlinks,linkcolor=blue,citecolor=blue,urlcolor=blue,breaklinks=true]{hyperref}

\DeclareMathOperator{\im}{Im}

\begin{document}

\title{Cavity optomechanics with arrays of thick dielectric membranes}

\author{Bhagya Nair}
\affiliation{Department of Physics and Astronomy, University of Aarhus, DK-8000 Aarhus C, Denmark}
\author{Andr\'e Xuereb}
\affiliation{Department of Physics, University of Malta, Msida MSD\,2080, Malta}
\affiliation{Centre for Theoretical Atomic, Molecular and Optical Physics, School of Mathematics and Physics, Queen's University Belfast, Belfast BT7\,1NN, United Kingdom}
\author{Aur\'elien Dantan}
\email[Corresponding author: ]{dantan@phys.au.dk}
\affiliation{Department of Physics and Astronomy, University of Aarhus, DK-8000 Aarhus C, Denmark}

\date{\today}

\begin{abstract}
Optomechanical arrays made of structured flexible dielectrics are a promising system for exploring quantum and many-body optomechanical phenomena. We generalize investigations of the optomechanical properties of periodic arrays of one-dimensional scatterers in optical resonators to the case of vibrating membranes whose thickness is not necessarily small with respect to the optical wavelength of interest. The array optical transmission spectrum and its optomechanical coupling with a linear Fabry-Perot cavity field are investigated both analytically and numerically. 
\end{abstract}

\pacs{42.50.Wk,42.50.Ct,85.85.+j}

\maketitle

\section{Introduction}
The level of control over the motion of mechanical oscillators using electromagnetic fields has recently increased tremendously enabling, e.g., their operation in the quantum regime~\cite{Aspelmeyer2014}. Optomechanical arrays, in which multiple mechanical oscillators can be coupled to several electromagnetic fields, expand the range of possibilities offered by these systems for exploring fundamental quantum and many-body phenomena~\cite{Bhattacharya2008multiple,Hartmann2008,Heinrich2010,Heinrich2011,Massel2012,Xuereb2012,Seok2012,Akram2012,Ludwig2013,Xuereb2014,Chesi2014,Kemiktarak2014,Xuereb2015,Schmidt2015,Buchmann2015,Lee2015,Weiss2016,Spethmann2016}
and for information processing or sensing applications~\cite{Stannigel2010,Cammerer2011,Stannigel2012,Zhang2012,Bagheri2013,Bagci2014,Andrews2014,Zhang2015,Fang2016}. 

Thin, flexible membrane resonators represent an attractive platform in this respect. Indeed, the use of a flexible membrane oscillator in the field of an optical resonator allows for benefitting from high-quality mechanical and optical quality factors~\cite{Thompson2008,Wilson2009,Cammerer2011,Karuza2012,Kemiktarak2012NJP,Purdy2013,Bagci2014,Andrews2014,Kemiktarak2014,Reinhardt2016,Norte2016,Yilmaz2016}. While experiments have so far focused on the use of single resonators, the interaction between multiple membrane oscillators and cavity fields has been investigated theoretically, e.g., for entanglement generation and nonlinear quantum optics~\cite{Bhattacharya2008multiple,Hartmann2008,Ludwig2013,Seok2012,Kipf2014}, the enhancement of radiation pressure forces~\cite{Xuereb2012,Xuereb2013,Tignone2013,Chesi2014}, and the engineering of long-range optomechanical interactions and many-body phonon dynamics~\cite{Heinrich2010,Xuereb2014,Chen2014,Xuereb2015}.

For periodic arrays of such membranes a well-suited theoretical framework is provided by the transfer matrix formalism~\cite{Deutsch1995}. In this one-dimensional formalism each membrane is described by a transfer matrix relating the forward- and backward-propagating waves on each side of the membrane. The generic optomechanical properties of the combined system consisting of the membrane array and optical resonator can be extracted through the application of standard methods for multilayered systems~\cite{Xuereb2009,Xuereb2012,Xuereb2013}. The case of two membranes presents a system for which optomechanical coupling strengths may be obtained analytically in a straightforward manner~\cite{Li2016}, and it is in fact the focus of this article. When performing such calculations, a convenient approximation, which is not \emph{a priori} necessarily justified in practice, consists in modelling the membranes by a beamsplitter whose thickness is much smaller than the field wavelength and which is characterized by its reflectivity (or, equivalently, polarizability). In doing so, one ignores phase shifts due to the propagation inside the dielectric and, for slabs which are thick enough compared to the wavelengths of interest, possible internal resonance effects due to multiple field oscillations within a slab. Evaluating these effects is thus highly relevant for practical implementations with membranes for which the thin-membrane approximation is typically not well met.

We address here these issues by investigating the effect of the membrane thickness on the transmission spectrum of a periodic array of flexible membrane resonators, as well as on the collective optomechanical coupling of the membranes with the field of an optical resonator. Based on a full transfer matrix approach we first show in Sec.~\ref{sec:optical} that arbitrarily thick membranes can be modelled as effective thin membranes and compute the transmission spectrum of a two-membrane array. In Sec.~\ref{sec:coupling} we investigate the optomechanical properties of such an array positioned at the center of a large optical resonator. We compute in particular the strength of the optomechanical couplings at specific ``transmissive'' wavelengths, where the array is effectively transparent and for which the field dispersively couples linearly to a collective motion of the individual membranes. We also make the connection with the results of Refs.~\cite{Xuereb2012,Xuereb2013}, obtained in the thin-membrane approximation, and extend them to the case when multiple field oscillations can occur \emph{inside} the individual membranes. We conclude in Sec.~\ref{sec:conclusion} and point out possible applications of these results.

\begin{figure*}[t]
\centering
\includegraphics[width=1.6\columnwidth]{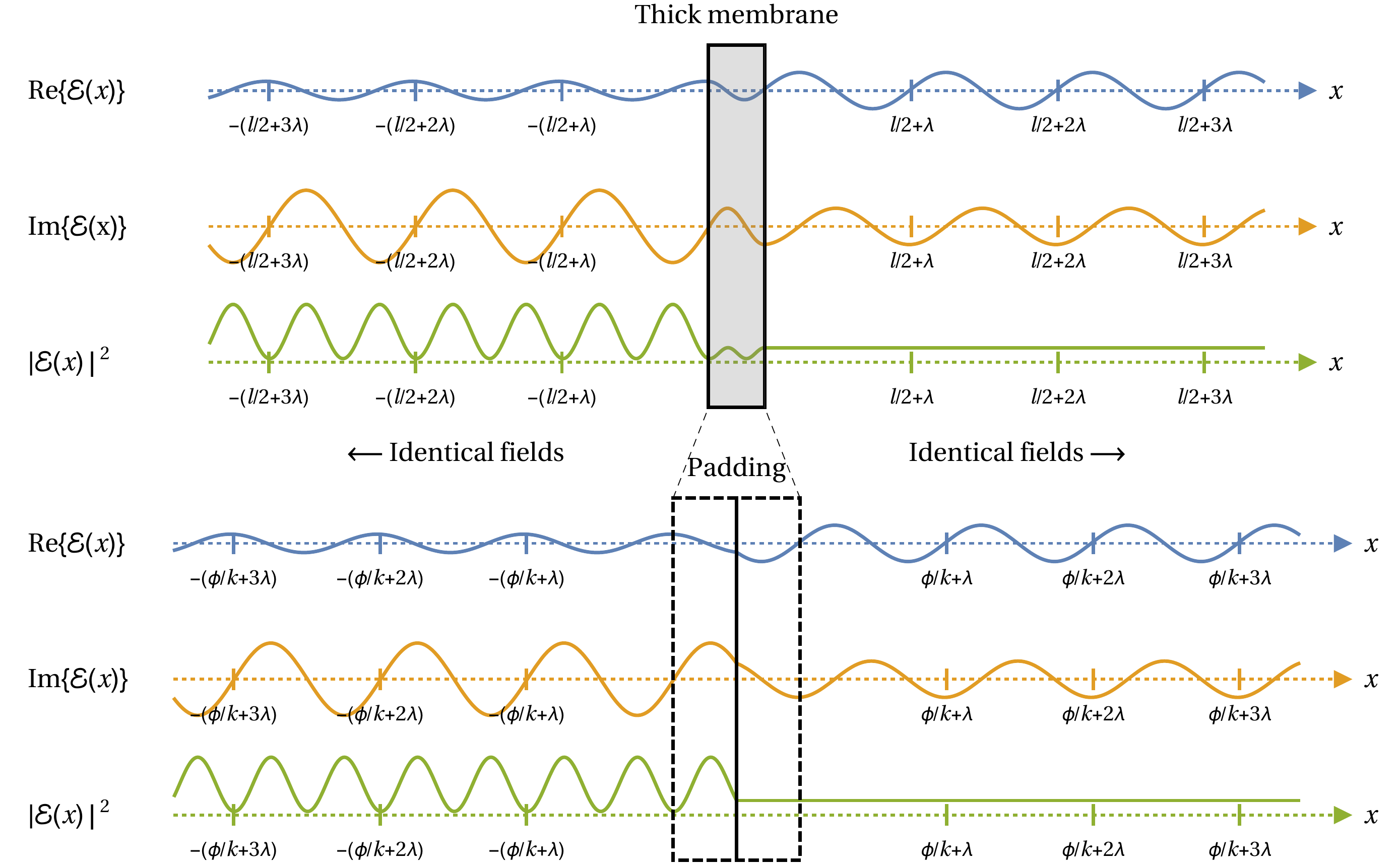}
\caption{Equivalence between a thick membrane (top) and an effective thin membrane with padding (bottom). From top to bottom in each of the subfigures, the curves illustrate the real (blue), imaginary (orange), and squared modulus (green) of the electric field, in arbitrary units, as a function of position. The shaded rectangle in the top subfigure shows the dielectric slab, while the dashed rectangle in the bottom subfigure show the extent of the padded areas around the infinitely thin membrane. Note that the fields outside the shaded (top) and dashed (bottom) rectangles are identical in both amplitude and phase.} \label{fig:equivalence}
\end{figure*}

\section{Optical properties}\label{sec:optical}

\subsection{Transfer matrix model}\label{sec:model}

As in previous studies~\cite{Xuereb2012,Xuereb2013} we restrict ourselves to one-dimensional systems and make use of the transfer matrix formalism, which is well-suited to model a periodic $N$-element array. In this formalism each element is described by a transfer matrix $M$ relating the forward- and backward-propagating waves on each side of a given element~\cite{Deutsch1995,Xuereb2009}
\begin{equation}
\biggl(\begin{array}{c}A\\B\end{array}\biggr)=M\biggl(\begin{array}{c}C\\D\end{array}\biggr)=\biggl[\begin{array}{cc}m_{1,1} & m_{1,2}\\ m_{2,1} & m_{2,2}\end{array}\biggr]\biggl(\begin{array}{c}C\\D\end{array}\biggr),
\end{equation}
with $A$ and $C$ ($B$ and $D$) are the amplitudes of the backward-propagating (forward-propagating) waves. For example, the free-space propagation of a monochromatic field of wavelength $\lambda=2\pi/k$ over a distance $d$ is described by the matrix
\begin{equation}
M_\text{fs}(d)=\biggl[\begin{array}{cc}e^{ikd} & 0\\0 & e^{-ikd}\end{array}\biggr].
\end{equation}
For an incoming field propagating to the right the transmissivity and reflectivity of the optical system modelled by $M$ are defined by
\begin{equation}
t=\frac{1}{m_{2,2}},\ \text{and}\ r=\frac{m_{1,2}}{m_{2,2}}.
\end{equation}

\subsubsection{Single membrane transfer matrix}
Each membrane is modelled as a dielectric slab with thickness $l$ and refractive index $n$. To simplify the discussion we assume the refractive index to be wavelength-independent and neglect absorption in the wavelength range considered, but these effects could easily be incorporated into our model. The Fresnel coefficients at normal incidence at the left and right vacuum--dielectric interfaces yield amplitude reflection and transmission coefficients
\begin{equation}
\rho_\text{l}=-\rho_\text{r}=\frac{1-n}{1+n}\equiv\rho,
\end{equation}
and
\begin{equation}
\tau_\text{l}=\frac{2}{1+n}\ \text{and}\ \tau_\text{r}=\frac{2n}{1+n},
\end{equation}
respectively. The transfer matrix of the slab with length $l$ can thus be written as
\begin{equation}
M_\text{m}=M_\text{l}M_\text{fs}(nl)M_\text{r},
\label{eq:Mm}
\end{equation}
where
\begin{equation}
M_{i}=\frac{1}{\tau_i}\biggl[\begin{array}{cc}1 & \rho_i\\ \rho_i & 1\end{array}\biggr]\ (i=\text{l},\text{r}).
\end{equation}
The reflection and transmission coefficients of the membrane are then given by~\cite{Jayich2008,Xuereb2012}
\begin{equation}
r_\text{m}=\frac{\rho(1-e^{2iknl})}{1-\rho^2e^{2iknl}},\ \text{and}\ t_\text{m}=\frac{\tau_1\tau_2e^{iknl}}{1-\rho^2e^{2iknl}}.
\label{eq:rtslab}
\end{equation}
The equivalent membrane polarizability $\zeta\equiv -ir_\text{m}/t_\text{m}$ is then
\begin{equation}
\zeta=\frac{n^2-1}{2n}\sin(knl).
\label{eq:zeta}
\end{equation}
Eqs.~(\ref{eq:Mm})--(\ref{eq:zeta}) hold for any membrane thickness. However, in the spirit of Refs.~\cite{Deutsch1995,Xuereb2012}, it can be convenient to model the membrane as an infinitely thin scatterer with an effective transfer matrix
\begin{equation}
\tilde{M}_\text{m}=\biggl[\begin{array}{cc}1+i\zeta & i\zeta\\-i\zeta & 1-i\zeta\end{array}\biggr],
\label{eq:Mmprime}
\end{equation}
where $\zeta$ is given by Eq.~(\ref{eq:zeta}), which gives reflection and transmission coefficients having the same amplitude as that of the equivalent membrane having arbitrary thickness. However, the thin-membrane model ignores the phase shift acquired by the field propagating through the membrane, which may be relevant, e.g., for propagation in a multi-membrane array and for taking into account field resonances inside individual membranes.

\begin{figure}
\centering
\includegraphics[width=0.98\columnwidth]{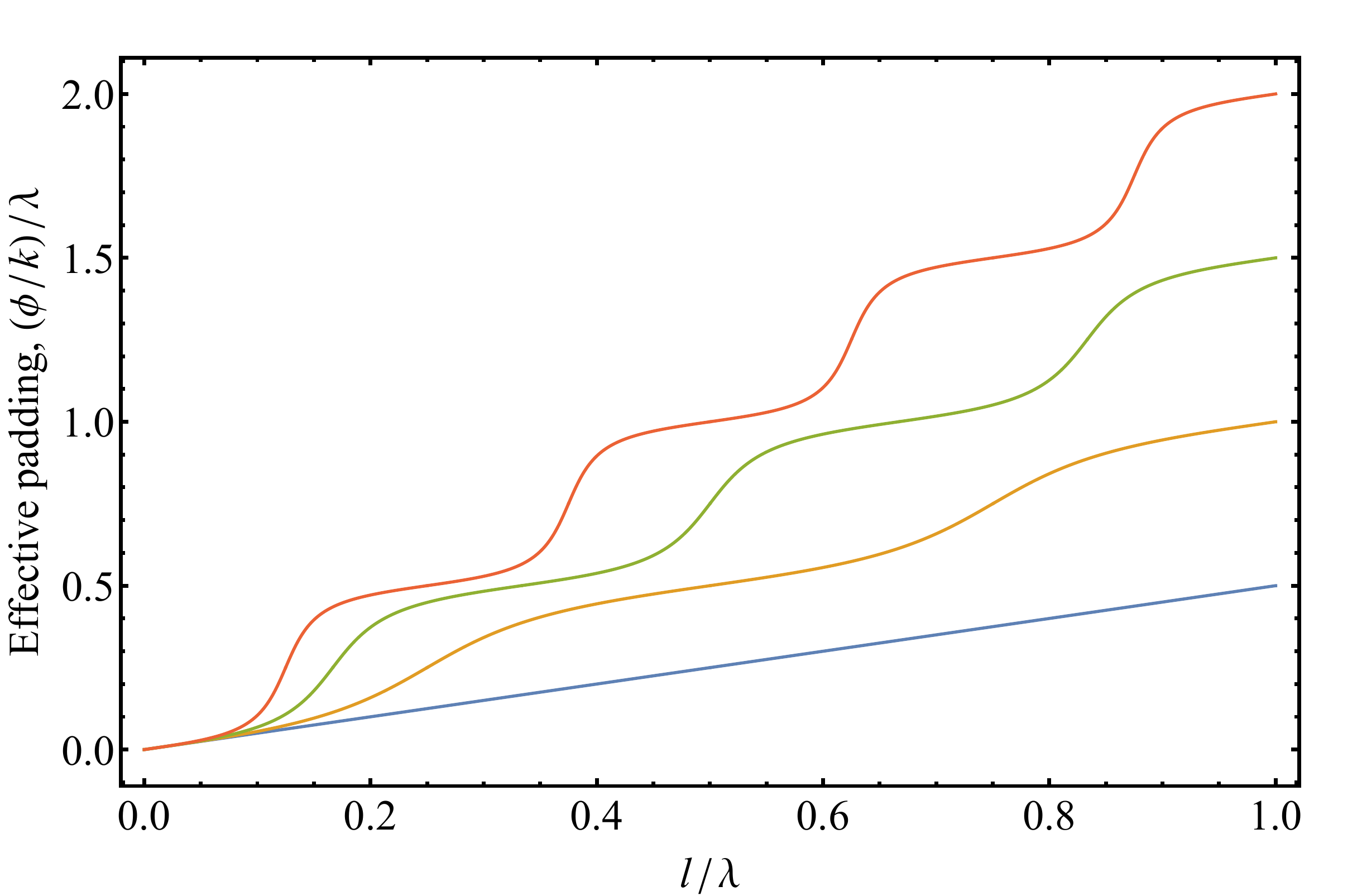}
\caption{Padding $\phi/k$, normalized to the wavelength $\lambda$, that is added to each side of the membrane in the thin-membrane model as a function of the thickness $l$ of the plate that the model is to match, shown for various values of the refractive index $n$. From bottom to top, the curves represent: $n=1$ (blue), $n=2$ (orange), $n=3$ (green), and $n=4$ (red).} \label{fig:padding}
\end{figure}

To take this phase shift into account one can introduce an extra padding of length $\phi/k$ to each side of the membrane so that its transfer matrix becomes 
\begin{equation}
M_\text{m}^{\prime}=M_\text{fs}(\phi/k)\tilde{M}_\text{m} M_\text{fs}(\phi/k),
\end{equation} with a padding phase
\begin{equation}
\phi=\begin{cases}
\phi_0+\pi\lfloor{nl/\lambda}\rfloor &\text{if}\ \sin(knl)>0\ \text{and}\\
2\pi-\phi_0+\pi\lfloor{nl/\lambda}\rfloor &\text{if}\ \sin(knl)<0,\\
\end{cases}
\end{equation}
where $\lfloor\cdot\rfloor$ represents the floor function and
\begin{equation}
\phi_0=\frac{1}{2}\arccos\biggl[\frac{(n^2-1)+(n^2+1)\sin(knl)}{(n^2+1)+(n^2-1)\sin(knl)}\biggr].
\end{equation}
As can be seen from the example shown in Fig.~\ref{fig:equivalence}, the resulting effective thin membrane conveniently models the propagation of the field outside the membrane. Fig.~\ref{fig:padding} illustrates $\phi$ as a function of the two parameters that describe the membrane. An interesting question to consider, which however is beyond the scope of the present work, is whether membranes where $\phi$ depends strongly on the thickness $l$ exhibit stronger coupling of the cavity field to dilational modes of the membrane~\cite{Borkje2012}.

\subsubsection{Periodic membrane array transfer matrix}

We consider a periodic array of $N$ identical, arbitrarily thick membranes, each modelled by a transfer matrix $M_\text{m}$ and separated by a distance $d$. The transfer matrix of the array is then computed as
\begin{equation}
M_N=M_\text{m}M_\text{fs}(d)M_\text{m}\cdots M_\text{m},
\label{eq:MN}
\end{equation}
where $M_\text{m}$ appears $N$ times. The transmittance of the array $\mathcal{T}=1/|(m_N)_{2,2}|^2$ can
be compared to that of the corresponding array of effective thin membranes $\mathcal{T}^{\prime}=1/|(M_N^{\prime})_{2,2}|^2$, where $M_N^{\prime}$ is defined by substituting $M_\text{m}^{\prime}$ for $M_\text{m}$ in Eq.~(\ref{eq:MN}).

\subsection{Two- and four-membrane arrays}

In this section we focus on the case of a two- and four-membrane arrays and use as an example silicon nitride membranes as employed in various membrane-in-the-middle experiments~\cite{Thompson2008,Wilson2009,Cammerer2011,Karuza2012,Kemiktarak2012NJP,Purdy2013,Bagci2014,Andrews2014}. 

\begin{figure}
\centering
\includegraphics[width=0.98\columnwidth]{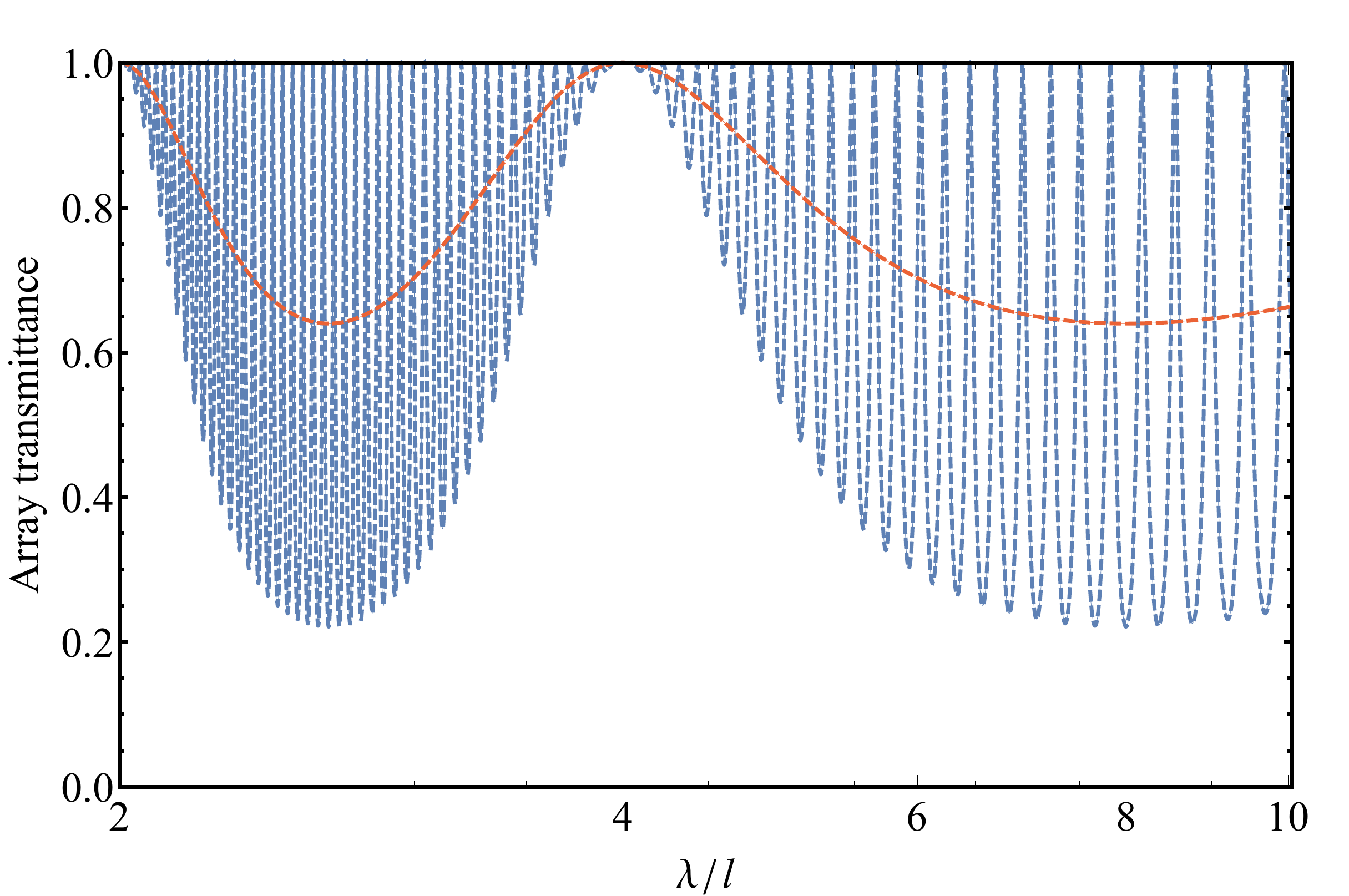}
\caption{The optical transmission spectrum of a two-membrane array with $n=2$ and $d=90l$. The full blue curve shows the transmittance ($\mathcal{T}$) resulting from the full transfer matrix calculations; superimposed on this curve is a dashed blue one that shows the transmittance ($\mathcal{T}^\prime$) from the effective thin-membrane model. The red curve shows the single-membrane transmittance ($|t_\text{m}|^2$) as a reference, with a dashed red curve superimposed on it calculated from the effective model. We note that the two models agree perfectly.} \label{fig:transmission}
\end{figure}

Figure~\ref{fig:transmission} shows the transmission spectrum of a two-membrane array with refractive index $n=2$, thickness $l=100$ nm and spacing $d=9$ $\mu$m, as experimentally investigated in~\cite{Nair2016}. The single-membrane transmittance spectrum is also displayed as reference. Unity transmission is achieved, as expected, when the reflectivity of the individual membrane, $r_\text{m}$, vanishes; this occurs when its effective thickness $nl$ is an integer multiple of $\lambda/2$. Unity transmission can also be achieved in a two-membrane array when there is perfect constructive two-mirror interference, which occurs at the ``transmissive" wavelengths discussed in Refs.~\cite{Xuereb2012,Xuereb2013}. As the figure illustrates, spectra resulting from the effective thin-membrane model (dashed curves) perfectly overlap with the results from the full model, showing the equivalence between the two models regarding free-space optical transmission.

\begin{figure}
\centering
\includegraphics[width=0.98\columnwidth]{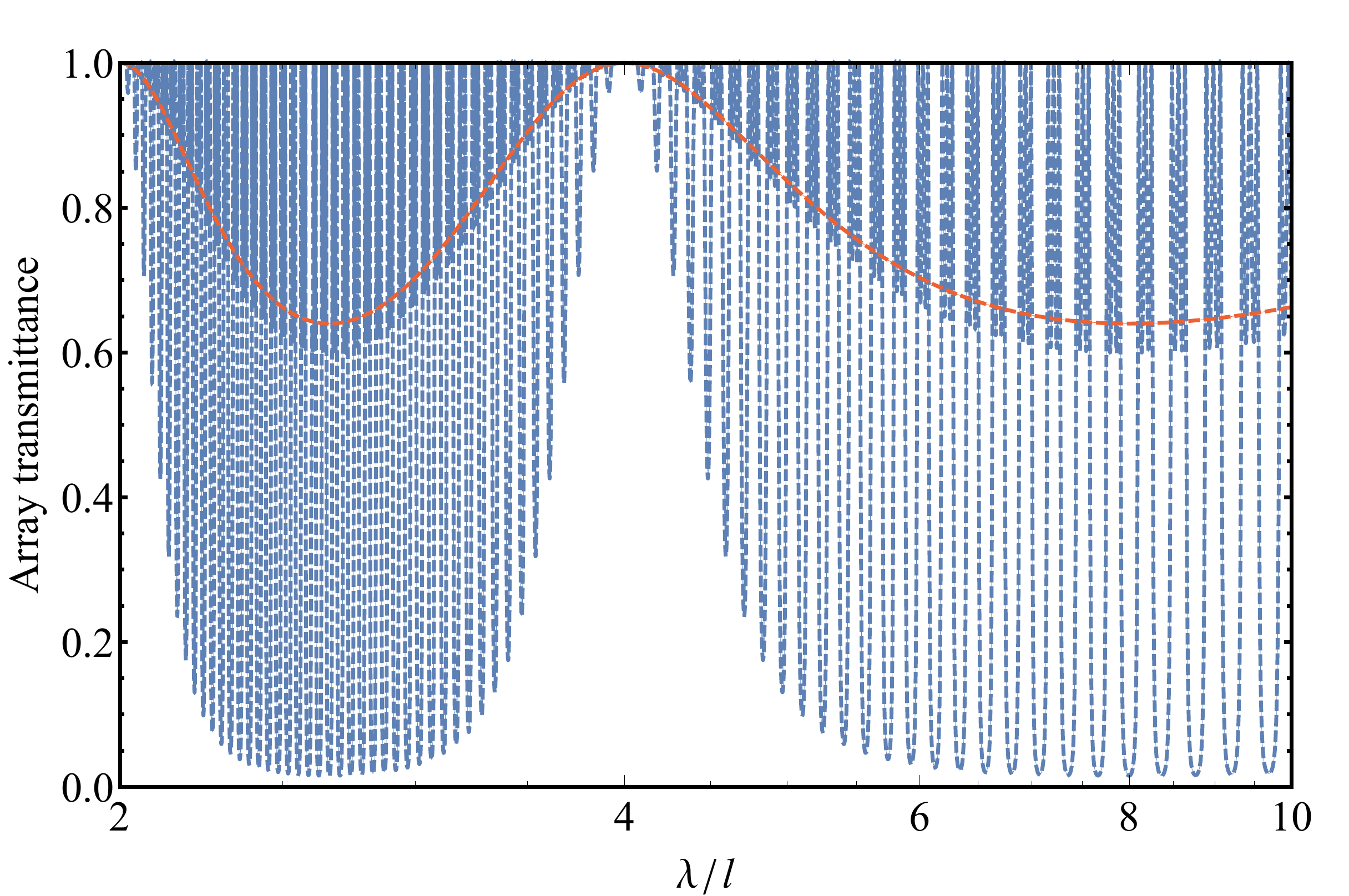}
\caption{Same as Fig.~\ref{fig:transmission}, but for a four-membrane array.} \label{fig:transmission4}
\end{figure}

Figure~\ref{fig:transmission4} shows the transmission spectrum of an array of four membranes with the same characteristics in the range $[3l,6l]$ around the first internal resonance wavelength. The overall interference pattern is similar to the two-membrane case, albeit with fully constructive interferences now occurring for triplets of close-by wavelengths~\cite{Xuereb2013}.

\section{Cavity optomechanics}\label{sec:coupling}

We now turn to the case of vibrating membranes positioned inside a (large) optical resonator and wish to investigate the role of the membrane thickness on the optomechanical coupling with the cavity field. Of particular interest, in connection with the results of Refs.~\cite{Xuereb2012,Xuereb2013}, is the strength of the optomechanical coupling at the aforementioned transmissive wavelengths, where the field couples dispersively and linearly to a collective motion of the individual membranes.

\subsection{Optomechanical coupling}

The $N$-membrane array, where $N=2$ or $4$ in the present paper, but our discussion applies generally, is assumed to be at the center of a symmetric linear Fabry--P\'erot cavity of length $L$. The length of the array is supposed to be much smaller than that of the cavity and the cavity field Rayleigh range. The cavity mirrors are modelled by a transfer matrix $M_\text{c}$ of the form (\ref{eq:Mmprime}) and their polarizability is denoted by $\zeta_\text{c}$. The total transfer matrix of the system can thus be written as the product
\begin{equation}
M_\text{tot}(L)=M_\text{c}M_\text{fs}(L_-)M_NM_\text{fs}(L_+)M_\text{c},
\end{equation}
where $L_{\mp}$ are the lengths of the sub-cavities to the left and the right of the array, respectively. Assuming the field wavelength to be equal to one of the transmissive wavelengths defined previously, it is easy to compute the cavity transmission spectrum as a function of $L$ in order to find the cavity resonances. In order to calculate the optomechanical coupling strength we follow the same method as in Refs.~\cite{Xuereb2012,Xuereb2013}: (i)~The cavity resonance frequencies $\omega$ are calculated for all membranes at their equilibrium positions, (ii)~the $j$\textsuperscript{th} membrane is then displaced by $\delta x_j$, (iii)~the corresponding transfer matrix calculated and the shift in the cavity resonances is computed, finally (iv)~yielding the individual optomechanical coupling $g_j$ of the $j$\textsuperscript{th} membrane through the relation $\omega\rightarrow\omega+g_j\delta x_j$. These coupling strengths define the collective motional mode of the membranes which is coupled to the field with a collective coupling constant
\begin{equation}
g_\text{coll}=\sqrt{\sum_{j=1}^N g_j^2}.
\end{equation}
As a figure of merit, $g_\text{coll}$ can be compared to the maximal coupling for a single perfectly reflective membrane, $g=2(\omega/L)x_\text{zpm}$, where $x_\text{zpm}$ is the extent of the wave-packet of the equivalent quantum harmonic oscillator in its ground state.

\subsection{Two-membrane array}

\begin{figure*}[t]
\centering
\includegraphics[width=2\columnwidth]{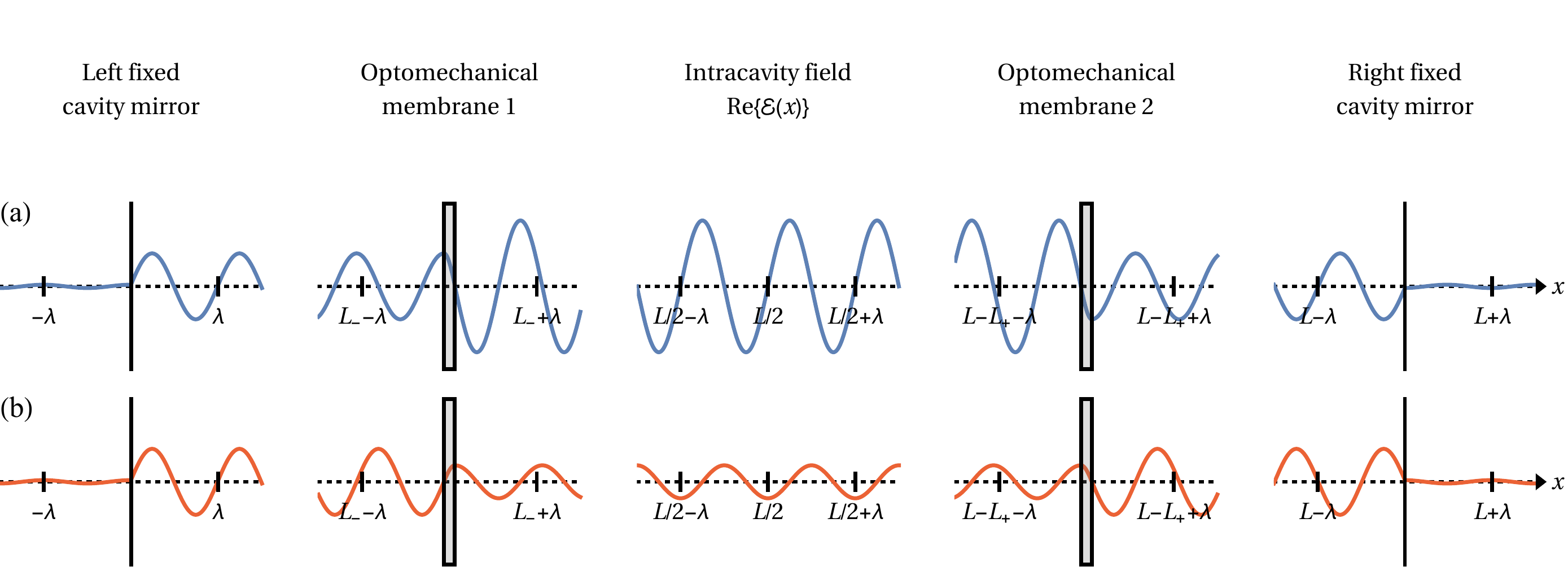}\\
\includegraphics[width=2\columnwidth]{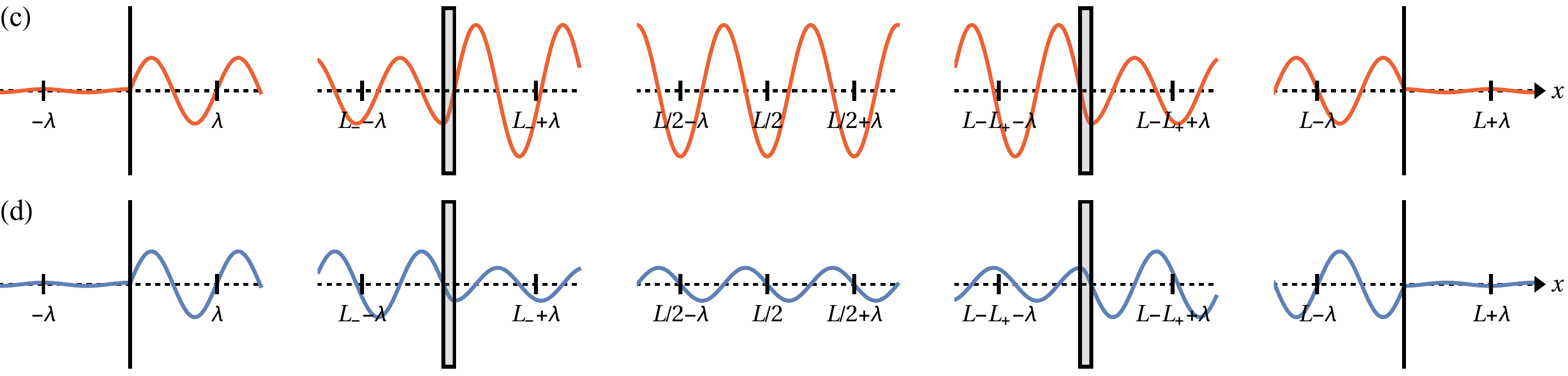}
\caption{Real part of the electric field (in arbitrary units) as a function of position for odd [blue; (a) and (c)] and even [red; (b) and (d)] cavity modes, for transmissive wavelengths larger than $nl$. From left to right, the five plots are centered around $x=0$, $L_-$, $L/2$, $L-L_+$, and $L$, respectively. The inner membranes have the same characteristics as those considered in Fig.~\ref{fig:transmission}, and the optical resonator has length $L\simeq 5\times 10^4l$ and finesse $3\,000$. The four parts of this plot correspond to the four labelled data points in Fig.~\ref{fig:OMcoupling}.} \label{fig:field}
\end{figure*}

We consider the case $N=2$ and assume that the field wavelength corresponds to one of the transmissive wavelengths, as in, e.g., Fig.~\ref{fig:transmission}. The shifts in the cavity resonance frequencies when one of the membranes is displaced by a small amount can then be calculated analytically in the same fashion as in Ref.~\cite{Xuereb2013}. One finds that the displacements of the membranes give rise to two different frequency shifts, which depend on the parity of the cavity mode number.  Figure~\ref{fig:field} shows as an example the real part of the electric field amplitude inside the cavity with the membranes are their equilibrium positions, for the case of two odd and two even cavity modes, and in the case $\lambda>2nl$ (no internal resonance). Cavity modes come in pairs; for each odd (even) cavity mode where the field amplitude between the membranes is increased as compared to its amplitude in the left and right subcavities, there exists an even (odd) cavity mode where the field amplitude is suppressed. The magnitude of the optomechanical coupling strength mimics the amplitude of the field between the membranes, i.e., it is larger in the case of the former set of modes and smaller in the latter case. In both cases, however, the resonance shifts are opposite for each membrane, which means that $g_1=-g_2$ and that the field couples to a breathing mode of the two membranes.

\subsubsection{Thin-membrane model: Optomechanical coupling strength}

To derive analytical expressions for the optomechanical couplings at the transmissive wavelengths of a two-membrane array we make use of the thin-membrane model, for which we are able to carry out analytical calculations. In the next section, we will compare the results obtained by replacing the polarizability that appears in the analytical coupling strengths obtained by using the thin-membrane approximation in this section by its general expression, which is valid for arbitrary membrane thickness.

The derivation of the ``transmissive" optomechanical couplings in the thin-membrane approximation proceeds along the same steps as in Sec.\ IIC of Ref.~\cite{Xuereb2013} and we only give the main steps here. Within the thin-membrane model the effective polarizability of the array can be shown to be
\begin{equation}
\chi=2\zeta(\cos\nu-\zeta\sin\nu),
\label{eq:chi}
\end{equation}
with $\nu=kd$. The array is transmissive when $\chi=0$, i.e., when
\begin{equation}
\cos\nu_{\pm}=\frac{\mp\zeta}{\sqrt{1+\zeta^2}}.
\label{eq:cosnu}
\end{equation}
The cavity resonance frequency shift for displacements $\delta x_1$ and $\delta x_2$ of membranes $1$ and $2$, respectively, is
\begin{equation}
\delta\omega=c\biggl(\frac{\partial k}{\partial \delta x_1}\delta x_1+\frac{\partial k}{\partial\delta x_2}\delta x_2\biggr),
\end{equation}
where $c$ is the speed of light in vacuum and the partial derivatives are given by Eqs.~(27)--(29) of Ref.~\cite{Xuereb2013}, i.e.,
\begin{equation}
\frac{\partial k}{\partial \delta x_1}=-\frac{\partial k}{\partial \delta x_2}=-\frac{\im\bigl\{\beta-e^{i\nu}\alpha\bigr\}}{L+2d\frac{\partial\chi}{\partial\nu}},
\end{equation} 
with $\alpha=2ik\zeta^2e^{-i\nu}$ and $\beta=-2k\zeta(1-i\zeta)e^{-i\nu}$. Using Eqs.~(\ref{eq:chi}) and~(\ref{eq:cosnu}), one gets that
\begin{equation}
\frac{\partial k}{\partial \delta x_1}=-2k\frac{\zeta(\pm\sqrt{1+\zeta^2}+\zeta)}{L[1\pm 4(d/L)\zeta\sqrt{1+\zeta^2}]}.
\end{equation}
Making further use of the fact that the resonance frequency shift is related to the normalized collective displacement by 
\begin{equation}
\delta\omega=g_{\pm}\frac{\delta x_1-\delta x_2}{\sqrt{2}},
\end{equation}
one obtains the collective optomechanical couplings
\begin{equation}
g_{\pm}=g\sqrt{2}\frac{\zeta(\pm\sqrt{1+\zeta^2}+\zeta)}{1\pm 4(d/L)\zeta\sqrt{1+\zeta^2}}.
\label{eq:gpm}
\end{equation}

The coupling $g_+$ is thus found to be identical to the one given by Eq.~(38) of Ref.~\cite{Xuereb2013}, albeit with a different sign convention for $\zeta$. We find $g_+$ to be larger than $g_-$ when the wavelength is large enough, $\lambda>2nl$, so that there is no internal resonance for the field inside a single membrane. However, in the region containing the first internal resonance (i.e., $nl<\lambda<2nl$), $g_-$ becomes larger than $g_+$. This can be understood by looking at the evolution of the intracavity field amplitude, as shown in Fig.~\ref{fig:field} in the case $\lambda>2nl$. For $\lambda>2nl$ cavity modes corresponding to $g_+$ show a greater field build-up between the membranes than the ones corresponding to $g_-$. The resulting radiation pressure forces and, therefore, the optomechanical coupling strength, are therefore stronger for modes corresponding to $g_+$. In contrast, for wavelengths such that $nl<\lambda<2nl$, because of the change in the sign of $\zeta$, the solution corresponding to cavity modes with a larger field build-up is found to be $g_-$. As $\lambda/l$ becomes smaller still, $g_+$ and $g_-$ alternate in a manner similar to the one just described.

\begin{figure}
\centering
\includegraphics[width=0.98\columnwidth]{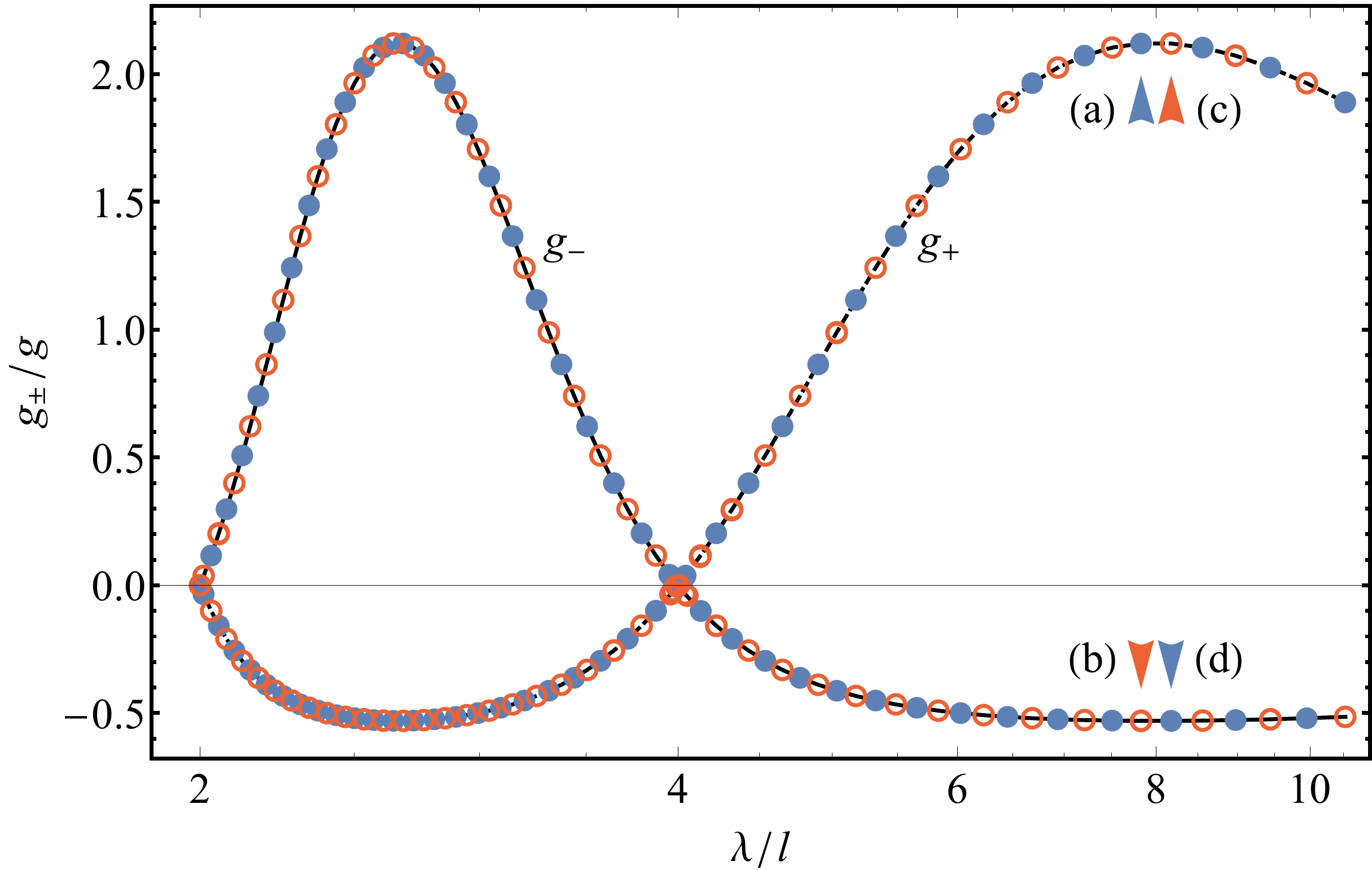}
\caption{Normalized optomechanical couplings $g_{\pm}/g$, at transmissive wavelengths for the same two-membrane array as used in Fig.~\ref{fig:transmission}. The array is positioned at the center of an optical resonator with length $5\times 10^4l$ and finesse $3\,000$. The data points show the results of the full transfer matrix calculations (solid blue circles for odd cavity modes, open red circles for even ones), while the curves show the results of the predictions based on the thin-membrane model [Eq.~(\ref{eq:gpm})]. Four resonances, labelled (a) through to (d), correspond to the respective field patterns shown in Fig.~\ref{fig:field}.} \label{fig:OMcoupling}
\end{figure}

It is interesting to consider the limiting cases for highly reflective membranes. For large $\lvert\zeta\rvert$ and $\lambda>2nl$,
\begin{equation}
g_+\sim g\frac{2\sqrt{2}\zeta^2}{1+4(d/L)\zeta^2}.
\end{equation}
As noted in Ref.~\cite{Xuereb2012} the denominator represents the relative increase in the effective length of the large cavity with length $L$ due to the field build-up in the small cavity, bounded by the membranes, with length $d$. As long as $4d\zeta^2/L\ll 1$, the effective length of the large cavity is unchanged and the optomechanical coupling strength scales as the finesse of the small cavity, which is proportional to $\zeta^2$. When the membranes are reflective enough to effectively narrow the large cavity linewidth, the optomechanical coupling saturates and tends to a value proportional to $\omega/d$, determined by the small cavity bounded by the membranes. 

In contrast, under the same conditions, we find
\begin{equation}
g_-\sim -\frac{g}{\sqrt{2}}\frac{1}{1-4(d/L)\zeta^2}.
\end{equation}
When $4d\zeta^2/L\ll 1$, the radiation pressure force on each membrane is provided by the field in the adjacent subcavity. Although the field amplitude in the shorter subcavities is the same as it would be in the large cavity without membrane array, interference between the two coupled subcavities reduces the optomechanical coupling strength. When the membranes are reflective enough, the reduced field amplitude between the membranes is leads to a reduction in the effective cavity length, which in turn results in an effective broadening of the cavity linewidth; this is the opposite situation to the one described in detail in Ref.~\cite{Xuereb2013}.

\subsubsection{Full transfer matrix model: Numerical results}
To investigate if the predictions of the thin-membrane model hold for realistic membranes with arbitrary thickness we numerically computed these cavity optomechanical coupling strengths at the transmissive wavelengths for the two-membrane array of Fig.~\ref{fig:transmission} using the method described above. The length of the cavity is taken to be $L=5\times 10^4 l$ and its finesse $3\,000$ as an example. Figure~\ref{fig:OMcoupling} shows both optomechanical coupling strengths $g_{\pm}$, normalized by $g$, numerically computed at each transmissive wavelength between $2l$ and $10l$. In both cases, the effective thin-membrane model predictions are well-corroborated by the full transfer matrix calculations, which justifies the role of the polarizability as the relevant parameter for characterizing the optomechanical properties of the system. It is interesting that similar optomechanical coupling strengths can be obtained, regardless of whether the membranes are thin or, on the contrary, thick enough for the field to oscillate several times within the dielectric medium, assuming equal effective masses. From Fig.~\ref{fig:OMcoupling} it is also clear that, for wavelengths close to an internal resonance, the optomechanical coupling strength vanishes, as there is no field imbalance across the membranes.

We also checked numerically for a four-membrane array, such as used in Fig.~\ref{fig:transmission4}, that the optomechanical coupling strengths agree with those predicted in Refs.~\cite{Xuereb2012,Xuereb2013} for arrays of infinitely thin movable scatterers.

\section{Conclusion}\label{sec:conclusion}
The transmission spectra and linear collective cavity optomechanical couplings of a periodic array of flexible membranes have been derived on the basis of full transfer matrix calculations taking into account the thickness of the membranes. The results support the use of the thin-scatterer approximation, provided a suitable phase-shift padding is introduced, and stress the role of the polarizability as the relevant parameter to investigate the optomechanical properties of these arrays. In a similar fashion it could also be interesting to investigate the role of defects~\cite{Tignone2013} and of patterning~\cite{Kemiktarak2012NJP,Reinhardt2016,Norte2016}, the dynamics of such arrays in the context of doped optomechanics~\cite{Dantan2014}, and the role of optical resonances of the membranes themselves in the context of dilational optomechanics~\cite{Borkje2012}.

\section*{Acknowlegments}
This work was supported by the Danish Council for Independent Research, the Carlsberg foundation, the European Commission (FP7-ITN {\it CCQED}), and COST Action MP1209 ``Thermodynamics in the Quantum Regime.''

\bibliography{thick_membrane_array_bib}

\end{document}